ФЕДЕРАЛЬНОЕ ГОСУДАРСТВЕННОЕ АВТОНОМНОЕ ОБРАЗОВАТЕЛЬНОЕ УЧРЕЖДЕНИЕ ВЫСШЕГО ОБРАЗОВАНИЯ

НАЦИОНАЛЬНЫЙ ИССЛЕДОВАТЕЛЬСКИЙ УНИВЕРСИТЕТ

«ВЫСШАЯ ШКОЛА ЭКОНОМИКИ»

**Факультет экономических наук**

**Образовательная программа " Прикладная экономика"**

**Кофнов Андрей Вадимович**

*«Фискальная теория уровня цен и состоятельность ее критики: пересмотр на основе стохастической модели с непрерывным временем»*

Выпускная квалификационная работа - МАГИСТЕРСКАЯ ДИССЕРТАЦИЯ

по направлению подготовки 38.04.01 Экономика

| | |
|---|---|
| **Рецензент** | **Руководитель** |
| **Научный сотрудник** | **Профессор Департамента** |
| **НУЛ макроэкономического анализа** | **теоретической экономики** |
| **НИУ ВШЭ** | **НИУ ВШЭ** |
| **Соколова** | **Ph.D., к.э.н.** |
| **Анна Валерьевна** | **Пекарский** |
| | **Сергей Эдмундович** |

**Москва 2017**

# Оглавление

# 1. Введение

Одной из основных задач правительства является поддержание стабильного уровня цен. При выполнении данной задачи необходимо ответить на вопросы о том, как достичь этой стабильности и насколько необходимо обеспечение устойчивой динамики цен. Первый вопрос можно переформулировать как "Может ли правительство повлиять на уровень инфляции?". Стандартный ответ на этот вопрос (версия монетаристского толка) заключается в том, что центральный банк влияет на уровень инфляции через управление денежным предложением. Иначе говоря, центральный банк полностью привержен политике таргетирования инфляции. Одной из проблем данного подхода является зависимость желания домохозяйств держать определенное количество денег от их ожиданий касательно будущей инфляции. В итоге этого пробела возможно существование множества равновесных траекторий динамики уровня цен в зависимости от ожиданий домохозяйств.

Как результат, некоторые экономисты начали переосмысливать основополагающие концепции данной теории. Рушится идея о том, что независимость центрального банка является достаточным условием стабильности ценовой динамики, и появляется гипотеза, что на нее может повлиять не только монетарная политика, но также и фискальная политика. Данные рассуждения порождают закономерный вопрос: может ли правительство (фискальные власти), используя стандартные инструменты фискальной политики, как варьирование уровня цен, государственных расходов и объема суверенного долга, дополнить политику монетарных властей для достижения устойчивого уровня цен?

Фискальная теория уровня цен (в литературе используется аббревиатура FTPL – Fiscal Theory of the Price Level) возникла как попытка ответить на вопросы, с которыми традиционный монетаристский подход оказался бессилен. Монетаристы в процессе объяснения природы инфляции наделяют монетар-ную политику уникальным правом нести ответственность за это явление, не учитывая действия фискальных властей. Идея подхода базируется на предположении, что уровень инфляции объясняется скоростью эмиссии денег. Следовательно, устойчивый уровень инфляции обеспечивается монопольным контролем над денежным предложением центрального банка и проводимым им сдерживающей монетарной политикой.

Данные аргументы говорят о том, что центральный банк является важным участником формирования ценовой стабильности. Но является ли он все-таки единственным агентом, отвечающим за уровень цен? Данный вопрос поднимается в работе Томаса Саржента и Ниила Уоллеса (T. Sargent, N. Wallace, 1981), на который авторы дают отрицательный ответ. Имеется необходимость учитывать влияние фискальной стороны.

В рамках данных рассуждений возникает проблема изучения взаимодействия фискальных и монетарных властей. Можно привести два аргумента в доказательство существования данного взаимодействия. В первую очередь центральный банк и правительство стремятся достигнуть общих целей, среди которых фигурируют обеспечение устойчивого экономического роста, а также поддержание стабильного уровня цен. Кроме того, правительство и центральный банк взаимодействуют ввиду существования консолидированного бюджетного ограничения, связывающего их активности. Это ограничение говорит о том, что дефицит бюджета может быть профинансирован как накапливанием государственного долга, за предложение которого отвечают фискальные власти, как и денежной эмиссией (сеньоражем), ответственным за которую выступает центральный банк.

Данная работа посвящена вопросу выявления устойчивого уровня инфляции, и консолидированное бюджетное ограничение сыграет центральную роль в выявлении равновесия. В основе работы лежит полемика, возникшая между главным идеологом фискальной теории уровня цен Майклом Вудфор-дом[1], предложившим нестандартный подход к интерпретации данного ограничения, которое впоследствии назовут нерикардианским условием, и его критиком Уиллемом Буайтером[2]. Аргументы последнего о несостоятельности данной теории отсылаются к несовершенности предлагаемых моделей с математической точки зрения из-за возможности возникновения множества точек равновесия в переопределенной системе уравнений.

Автор новой теории вводит новый тип условия для получения нового условия равновесия, так как в монетарном подходе формирования моделей предусмотрена заведомая недоопределенность системы, неполнота определения равновесия. Новое условие доопределяет систему и позволяет получить равновесие единственным образом.

В данной работе будет показано, что вопрос об определенности или переопределенности модели является проблемой методологического характера. Возможно, что аргументы Буайтера построены на таком построении модели экономики в его работе, специфика которой и ведет к озвученным им и другими критиками проблемам. Далее будет предъявлена модель такой спецификации, в которой будет учтен ряд параметров, которыми оперировал Буайтер в процессе своей критики, и при этом будет получено единственное равновесие, где уровень цен определяется однозначно.

В следующей главе будет проведен обзор литературы, которая послужила базой для исследования в данной работе. Особое внимание будет уделено именно процессу моделирования для реализации идеи фискальной теории уровня цен, подробно рассмотрены спецификации моделей как автора теории, так и ее критика.

В третьей главе будет подробно описана структура специально разработанной для поиска компромисса между авторами и критиками теории моде-ли, представлены основные выкладки о решении задачи динамического программирования, выведено условие равновесия. Модель данной работы будет реализована в условиях непрерывного времени и стохастического характера протекающей динамики. Поэтому в начале третьей главы будут представлены основные теоретические элементы стохастического моделирования с непрерывным временем. Большинство экономистов, включая критиков теории, строят свои иллюстрации экономических процессов в дискретном времени. В силу этого архитектура моделей с непрерывным временем представляет особый интерес, так как не было выявлено аналогичных работ известных авторов, кто бы практиковал применение такого рода моделирование в рамках фискальной теории определения уровня цен.

---

[1] См. Woodford M. Price-level determinacy without control of a monetary aggregate. // Carnegie-Rochester Conference Series on Public Policy 43, 1995, p. 1-46.
[2] См. Buiter W. The fiscal theory of the price-level: a critique. // The Economic Journal, 112, July 2002, p. 459-480.

В заключении сформулированы основные результаты проделанной работы, а также идеи и перспективы дальнейших исследований.

## 2. Обзор литературы

Развитие фискальной теории уровня цен (FTPL) привнесло новые идеи для исследования в современной макроэкономической теории. Одним из центральных объектов для исследования становится взаимодействие фискальной и монетарной политик. Одной из фундаментальных работ в этой области можно выделить статью Роберта Лукаса и Нэнси Стокей (Lucas R.E. Jr., Stokey N.L., 1983), в которой тестируется возможность предъявить оптимальные фискальную и монетарную политики в экономике без капитала в терминах согласованности во времени в бартерной и монетарной моделях. Авторы заключили, что использование в монетарной модели проведения дискреционной политики монетарными властями для взимания "инфляционного налога" не может быть осуществлено связыванием его с величиной государственных обязательств, как это возможно сделать с налогами фискальных властей. Выявлено было, что согласованность во времени достигается только тогда, когда монетарная политика носит пассивный характер – настроена на поддержание определенной траектории уровня цен.

Интересна более ранняя статья Карла Крайста (Christ C.F., 1979), в которой автор утверждает о нецелесообразности выделения фискальной и монетарной политик в два отдельных элемента как действий двух независимых агентов. Аргумент Крайста сводится к тому, что при такой ситуации активные действия одного из агентов и соответствующе изменение какой-либо из политических переменных без компенсирующего воздействия другого агента может привести к нарушению консолидированного бюджетного ограничения. А, следовательно, рассмотрение такой ситуации не имеет смысла. Поэтому необходимо глобальное описание всей макроэкономической политики в целом с явным определением всех экзогенных переменных, отвечающих соответствующему действию одного из участников консолидированного агента, и эндогенных переменных, отображающих соответствующую реакцию на это изменение другого агента.

Существенным толчком к развитию фискальной теории уровня цен стала работа Нила Уоллеса и Томаса Саржента "Некоторая неприятная монетаристская арифметика" (Sargent T., Wallace N., 1981[3]). В данной работе авторы демонстрируют, что достижение целевого уровня инфляции не осуществить одними лишь усилиями монетарных властей, необходима способствующая активность фискальной стороны. Предложив свои модели как частный случай, авторы делают акцент на потребности учитывать влияние правительства на уровень цен, а не только центрального банка.

Авторы рассматривают два противоположных режима: активная монетарная политика – пассивная фискальная политика и наоборот: активная фискальная политика – пассивная монетарная. В первом случае, когда фискальная политика подстраивается под действия доминирующей монетарной политики, центральному банку удается контролировать инфляцию на длительный отрезок времени. Иначе обстоят дела со вторым режимом. Так как имеется консолидированное бюджетное ограничение, то при активной фискальной политике монетарным властям приходится обеспечивать образуемый долг необходимым объемом сеньоража для поддержания его устойчивости. И этот пример становится прецедентом для обсуждения проблемы невозможности поддержания устойчивого уровня инфляции одними лишь усилиями монетарных властей.

Фискальные и монетарные власти принимают решения касательно макроэкономической политики только в соответствии с консолидированным бюджетным ограничением. Таким образом, значения всех величин, описывающих политику, не могут быть экзогенными – как минимум одна из них обязана быть эндогенной, то есть один из политиков подстраивается так, чтобы выполнялось равенство в бюджетном ограничении.

---

[3] Thomas J. Sargent, Neil Wallace. Some unpleasant monetary arithmetics // Federal Reserve Bank of Minneapolis, Quaterly Review, Fall 1981, P.2.

Существуют другие интересные работы, посвященные различным режимам фискальных и монетарных политик. Майкл Вудфорд в своей работе (Woodford M., 1994) на основе монетарной экономики Лукаса и Стокей (Lucas R.E. Jr., Stokey N.L., 1983) тестирует два типа монетарных режимов: экзогенный уровень роста предложения денег и фиксированная процентная ставка по однопериодным государственным облигациям. Автор показал, что при экзогенном предложении денег возможны ситуации наличия множества точек равновесия и вырожденных равновесий (отрицательные значения тех переменных, которые не могу быть отрицательными по определению, таких как уровень цен). При фиксированной процентной ставке автор утверждает, что можно определить единственное равновесие в модели, и доказывает невозможность вырожденных равновесий.

Имеется ряд работ, в которых была задействована теория игр для реализации моделей взаимодействия правительства и центрального банка. Среди прочих можно выделить статьи Алана Блайндера (Blinder A., 1982), Альберто Алезина и Гвидо Табеллини (Alesina A., Tabellini G., 1987), Биитсма и Бовенберга (Beetsma R.M.J.W., Bovenberg L., 1995, 1997).

Фискальная теория уровня цен зародилась в начале 90-х гг. 20-го века и заручилась вниманием многих крупнейших ученых-экономистов современности. Статья Эрика Липера "Равновесия при "активных" и "пассивных" монетарных и фискальных политиках" (Leeper E.M., 1991) считается первой работой, которая стала прецедентом к зарождению фискальной теории определения уровня цен. В этой работе он рассматривал взаимодействия властей в стохастической модели общего равновесия с дискретным временем, где под терминами "активная" и "пассивная" понимал характер реакции политика на шок государственного долга. Липеру в его модели удалось выявить детерминированное равновесие только в двух возможных режимах из четырех: когда одна из политик активная, а другая – пассивная. Основными работами в данной области стали статьи Майкла Вудфорда (Woodford M., 1995, 1996, 1998, 2001), Кристофера Симса (Sims C.A., 1994) и Джона Кокрейна (Cochrane J.H., 2000, 2001, 2003).

Отличительной чертой данной теории является то, под каким углом стоит рассматривать консолидированное бюджетное ограничение правительства. Если в других более традиционных теориях бюджетное ограничение является незыблемым равенством и должно выполняться при любых значениях исследуемых переменных (так называемая рикардианская политика), то у FTPL свой собственный взгляд на это ограничение. Для фискальной теории уровня цен является допустимым невыполнение данного ограничения для некоторых значений переменных, и оно является очередным уравнением, под которое не надо подстраиваться при поиске равновесия. FTPL требует выполнение равенства в консолидированном ограничении лишь для равновесного состояния (равновесной траектории уровня цен), а, значит, это уравнение может выступать ориентиром, по которому можно определить в ходе решения системы уравнений находимся ли мы в равновесии или нет.

$$\frac{B(t)}{P(t)} = \int_t^{+\infty} \{s(z) - d(z)\}e^{-r(z-t)}dz \qquad (2.1)$$

В моделях фискальной теории уровня цен государственный долг рассматривается в номинальных величинах. Переход же к реальным величинам происходит корректировкой на текущий уровень цен. Получается, бюджетное ограничение выполняется, когда дисконтированная сумма ожидаемого сеньоража и профицита бюджета совпадет с накопленным реальным долгом к моменту времени $t$. Если равенство не выполняется, то уровень цен находится не на равновесной траектории, он становится переменной, значение которой надо определить, при котором целевая функция – бюджетное ограничение – примет требуемое значение.

Этот прием используется при моделировании в задачах на экстремум, когда у домохозяйств определяется функция полезности, их бюджетное ограничение, а консолидированное ограничение государственного сектора пред-ставляет собой аналог целевой функции, дающее очередное уравнение для поиска равновесия.

В работе "Price-level determinacy without control of a monetary aggregate" (Woodford M., 1995[4]) Вудфорд показал, что уровень цен в экономике может быть определен даже при эндогенности предложения денег. Такое возможно при условиях, которые он обобщил термином "фискальная теория уровня цен", в соответствии с которыми равновесная траектория динамики цен есть такая траектория, при которой реальное значение выраженных в номинальных величинах государственных обязательств совпадает с дисконтированным значением ожидаемых будущих излишков.

Рассмотрим модель, которую автор предложил для решения вынесенной в описание статьи проблемы и посмотрим, к каким результатам он пришел.

Автор предложил для рассмотрения модель монетарной экономики для репрезентативного домохозяйства с функцией полезности, построенной по аналогии с вариантом, предложенным Мигелем Сидрауски (Sidrauski M., 1967), где функция полезности зависит от реального потребления $c_t$ и реальных денежных балансов $m_t = M_t/p_t$, где $p_t$ обозначает уровень цен, а $M_t$ - объем денежной массы.

$$U_t = U_t(c_t, m_t) \tag{2.2}$$

Функция является вогнутой и возрастает по обоим аргументам. Задача домохозяйств сводится к максимизации суммы дисконтированных полезностей:

$$\sum_{t=0}^{\infty} \rho^t U_t(c_t, m_t), \tag{2.3}$$

где $0 < \rho < 1$ - дисконтирующий множитель. Отнесение денежных балансов к аргументам функции полезности соотносится с идеей, что домохозяйствам нужен инструмент, обеспечивающий проведение транзакций, а также являет-ся чисто техническим элементом для определения спроса на деньги. Вудфорд интерпретирует $M_t$ как объем денежной базы, что объединяет в себе наличность, находящуюся на руках у домохозяйств, и депозиты у финансового сектора, которая в явном виде в модели не присутствует.

---

[4] Woodford M., 1995, Ibid.

Модель Вудфорда полностью детерминированная, поведения монетарных и фискальных властей полностью определены, а равновесие ищется при совершенном предвидении.

Финансовый рынок в этой представляет себя выбор между всего двумя активами – деньгами и однопериодными государственными облигациями. И деньги, и облигации наделены доходностью. Доходность денег в момент времени $t$ обозначается $R_t^m$, а доходность облигаций обозначается как $R_t^b$.

Максимизация функции (2.3) производится при бюджетном ограничении (2.4)

$$p_t c_t + M_t + B_t \leq W_t + p_t y_t - T_t, \qquad (2.4)$$

где $W_t$ обозначает номинальный объем богатства на начало периода, $B_t$ и $M_t$ обозначают номинальную стоимость государственных облигаций и объем денежной массы соответственно на в собственности домохозяйств на конец периода, $T_t$ – номинальный объем чистых (с вычетом трансфертов) аккордных налогов, выплаченных за период $t$, $y_t$ - реальный доход (количество потребительских товаров, которым наделяется домохозяйство) на период времени $t$.

Богатство домохяйств на следующий период времени определяется по закону (2.5):

$$W_{t+1} = M_t R_t^m + B_t R_t^b. \qquad (2.5)$$

Домохозяйства в каждый момент времени принимают решение об объеме потребления и сбережений, а точнее о структуре своего инвестиционного портфеля: $c_t, M_t \geq 0$. Вудфорд не накладывает ограничения – требования об неотрицательности объемов государственных облигаций $B_t$, а значит, домохозяйства могут как давать в долг государству, так и сами занимать деньги. Но Вудфорд предъявляет ограничение по объему займов для домохозяйств. Каждый новый момент времени для домохозяйств должен начинаться с объемом богатства не меньшим, чем указано в уравнении (2.6).

$$W_t \geq -\sum_{j=0}^{\infty} \frac{p_{t+j} y_{t+j} - T_{t+j}}{\prod_{s=0}^{j-1} R_{t+s}^b} \qquad (2.6)$$

Данное ограничение означает, что для домохозяйств допустим любой долг, который может быть выплачен за конечное число интервалов времени. Вудфорд отмечает, что необходимо накладывать определенные ограничения на объемы займов домохозяйств для четкого определения бюджетного множества. Иначе возможны возникновения ситуаций с "пирамидами" Понци.

Далее домохозяйства выбирают тройку потребление - денежная масса – номинальный объем облигаций $\{c_t, M_t, B_t\}$ при условиях (2.4) – (2.5) и ограничении (2.6) для всех моментов времени $t \geq 0$ при заданном начальном значении объема богатства $W_0$, траекториях уровня цен и процентных ставок $\{p_t, R_t^b, R_t^m\}$ и траекториях реального дохода и уровня налогов $\{y_t, T_t\}$. Эти последовательности описывают равновесие при совершенном предвидении, когда спрос на денежные балансы уравновешивается соответствующим предложением денег $M_t$, спрос на государственные облигации $B_t$ совпадает с объемом, выпускаемым государством, и выполняется макроэкономическое тождество

$$c_t + g_t = y_t, \qquad (2.7)$$

где $g_t$ – реальный объем государственных расходов за период $t$.

Режим государственной фискальной и монетарной политики определятся четырьмя переменными из следующих шести: $\{g_t, T_t, R_t^m, R_t^b, M_t, B_t\}$, где каждый из этих элементов может быт функцией от уровня цен $p_t$. При этом только четыре из шести переменных могут быть независимо выбраны политиком. Это необходимо, так как другие переменные подстраиваются под сделанный выбор, а в каждый момент времени должно выполняться бюджетное ограничение

$$p_t g_t = T_t + (M_t - M_{t-1} R^m_{t-1}) + (B_t - B_{t-1} R^b_{t-1}). \qquad (2.8)$$

Кроме того Вудфорд отмечает, что государство не может одновременно экзогенно определять и цены на товары в денежном измерении, а также их количества. Это означает, что если государство определяет объем выпускаемых облигаций и денег, то их цены (их доходности) должны быть определены в равновесии на рынке. А также наоборот: если государство устанавливает доходности денег и облигаций, то должно разрешить домохозяйствам при заданных величинах самим формировать свой инвестиционный портфель. Поэтому экзогенно может быть выбрано не более трех переменных из следующей четверки: $\{M_t, B_t, R^m_t, R^b_t\}$.

Вудфорд приводит пример, что в соответствии с идеями, выдвигаемыми в количественной теории денег, где предполагается экзогенный выбор динамики роста денежной массы $M_t$ и отсутствие государственных займов, можно определить спецификацию с экзогенным заданием набора последовательностей $\{g_t, B_t, R^m_t, M_t\}$. Данная спецификация обычно подразумевает, что $R^m_t = 1$, а $B_t = 0$ для всех $t$. Тогда траектория налогов $T_t$ определяется из уравнения (2.8), а доходность облигаций $R^b_t$ выявляется из равновесия на рынке ценных бумаг.

Вудфорд в статье рассматривает два альтернативных режима. В первом он экзогенно задает тройку $\{g_t, R^m_t, M_t\}$, а уровень налогов $T_t$ определяется по правилу, предусматривающему эндогенное определение набора $\{p_t, R^b_t\}$, а объем выпуска государственных облигаций $B_t$ выявляется через (2.8). В качестве другого режима Вудфорд рассматривает управление процентной ставкой через экзогенное задание доходностей $\{R^m_t, R^b_t\}$, а в это время структура государственных обязательств определяется через равновесие на рынке. Если фискальные власти определят траекторию динамики государственных закупок и объема налогов, то объем денежной массы и облигаций определяются через уравнение (2.8).

Сформулируем условие равновесия в модели при полном предвидении. Заметим, что условия (2.4) и (2.8) равносильны следующему неравенству

$$\sum_{t=0}^{\infty} \frac{p_t c_t + \Delta_t M_t}{\prod_{s=0}^{t-1} R_s^b} \leq W_0 + \sum_{t=0}^{\infty} \frac{p_t y_t - T_t}{\prod_{s=0}^{t-1} R_s^b}, \qquad (2.9)$$

представляющее собой бюджетное ограничение в текущих значениях при заданном начальном богатстве $W_0$, где

$$\Delta_t = \frac{R_t^b - R_t^m}{R_t^b}, \qquad (2.10)$$

а запланированные объемы потребления и денег на руках на все периоды времени $t \geq 0$ удовлетворяют условию о неотрицательности. Функцию (2.10) можно интерпретировать как "издержки" держания денег, то есть предельную норму замещения (так как недополучаются процентные платежи по облигациям). Тогда запланированные объемы потребления и денежной массы должны удовлетворять следующим условиям первого порядка:

$$\frac{u_m(c_t, m_t)}{u_c(c_t, m_t)} = \Delta_t, \qquad (2.11)$$

$$u_c(c_t, m_t) = \rho(1 + r_t^b) u_c(c_{t+1}, m_{t+1}), \qquad (2.12)$$

которые должны выполняться для всех $t \geq 0$, где $m_t = M_t/p_t$ отображают реальные денежные балансы, $r_t^b$ - реальная ставка доходности по облигациям,

$$r_t^b = R_t^b \left(\frac{p_t}{p_{t+1}}\right) - 1, \qquad (2.13)$$

а условие (2.9) обращается в равенство. Очевидно, что должны существовать такие оптимальные планы $\{c_t, M_t\}$, при которых левая и правая часть условия (2.9) ограничены. Иначе пропадает ограничение на уровень потребления, а в этом случае оптимального потребления не существует.

Если и потребление, и реальные денежные балансы являются нормальными благами, то соотношение $u_m/u_c$ возрастает с ростом $c$ и убывает с ростом $m$. этого следует, что (2.11) можно переписать в следующем виде:

$$m_t = L(c_t, \Delta_t), \quad (2.14)$$

где функция спроса на ликвидность $L$ возрастает с ростом $c$ и убывает с ростом $\Delta$. Подставляя (2.7) в (2.11), приходим к выражению, которое дает еще одно условие равновесия:

$$m_t = L(y_t - g_t, \Delta_t). \quad (2.15)$$

Это стандартное уравнение кривой LM, в котором спрос на деньги предполагается зависимым от объемов государственных закупок. Подставляя (2.7) в (2.12), приходим к еще одному условию

$$\lambda(y_t - g_t, \Delta_t) = \rho(1 + r_t^b)\lambda(y_{t+1} - g_{t+1}, \Delta_{t+1}), \quad (2.16)$$

где $\lambda(c, \Delta) = u_c(c, L(c, \Delta))$. Это оптимизированная версия хиксианской кривой IS. Подстановка (2.7) в (2.9) дает нам уравнение межвременного бюджетного ограничения

$$\frac{W_0}{p_0} = \sum_{t=0}^{\infty} \frac{(\tau_t - g_t) + \Delta_t m_t}{\prod_{s=0}^{t-1}(1 + r_s^b)}, \quad (2.17)$$

где $\tau_t = T_t/p_t$ отображает реальный объем налогов в момент времени $t$. Условие (2.17) означает, что текущее значение будущих излишков бюджета должно совпадать с текущим значением государственных обязательств. Данное условие должно выполняться не только для начального момента времени

$t = 0$, но и для всех последующих. Тогда (2.17) можно переписать в следующем виде:

$$\frac{W_t}{p_t} = \sum_{s=t}^{\infty} \frac{(\tau_s - g_s) + \Delta_s m_s}{\prod_{j=t}^{s-1}(1+r_j^b)}. \qquad (2.18)$$

Уравнение (2.18) можно рассматривать как условие равновесия, которое определяет уровень цен $p_t$ на период $t$ при заданных значениях чистых номинальных государственных обязательств $W_t$ и ожиданиях о текущих и будущих значений переменных, которые фигурируют в правой части уравнения (2.18). Величина $W_t$ определяется величиной предыдущих излишков бюджета из следующего уравнения динамики:

$$W_{t+1} = R_t^b [W_t + p_t(g_t - \tau_t - \Delta_t m_t)], \qquad (2.19)$$

которое выводит подстановкой (2.8) в (2.5).

Равновесие в модели описывается набором последовательностей $\{g_t, T_t, R_t^m, R_t^b, M_t, p_t\}$, который согласуется с фискально-монетарным режимом политики и обеспечивает ограниченность правой части неравенства (2.9), а также удовлетворяет уравнениям (2.15), (2.16) и (2.18) для всех $t \geq 0$ при заданном начальном богатстве $W_0$.

Заметим, что уравнение (2.18) равносильно уравнению (2.19) при выполнении условия трансверсальности:

$$\lim_{T \to \infty} \frac{W_T}{\prod_{s=0}^{T-1} R_s^b} = 0. \qquad (2.20)$$

Таким образом, равновесие в модели при полном предвидении может быть эквивалентно определено как набор последовательностей $\{g_t, T_t, R_t^m, R_t^b, M_t, p_t\}$, который согласуется с фискально-монетарным режимом политики и обеспечивает ограниченность правой части неравенства (2.9), а также удовлетворяет уравнениям (2.15), (2.16) и (2.19) для всех $t \geq 0$ и удовлетворяет условию (2.20) при заданном начальном богатстве $W_0$.

На основе данной модели Вудфорд получает следующий результат. Во-первых, определение уровня цен возможно даже при таких фискально-монетарных режимах, в которых предложение денег не экзогенно, а центральный банк управляет процентной ставкой. Во-вторых, выявление стабильности уровня цен эквивалентно изучению вопроса стабильности ожиданий о будущей политике правительства.

Вудфорд формулирует свои тезисы при ослаблении условия (2.18). Традиционно считается, что данное условие должно соблюдаться при любом уровне цен. Автор приходит к своим результатам при трактовке о выполнении (2.18) только лишь при равновесном уровне цен.

В то же время ряд экономистов критикуют данную теорию, апеллируя в основном некорректностью переформулировки условия (2.18). Так Дирк Нипельт в своей работе (Niepelt, D., 2004) утверждает, что предположение о не-рикардианском условии (переформулировка (2.18) по Вудфорду) и определение уровня цен инструментами фискальной политики не совместимы с равновесием, в котором активы домохозяйств должны отражать их оптимальный выбор. Он настаивает, что условие должно соответствовать рикардианской формулировке (выполняться при любом уровне цен) при таком определении равновесия.

В другой работе Нараяна Кочерлакота и Кристофер Фелан (Kocherlakota N., Phelan C., 1999) основательно подходят к разбору работы Вудфорда. Во-первых, они показали, что фискальная политика может оказать воздействие на уровень инфляции тогда и только тогда, если правительство ведет себя кардинально отличающимся от поведения домохозяйств образом. Если домохозяйства должны придерживаться своего бюджетного ограничения, невзирая на уровень цен, то, аналогично идее Вудфорда, правительство при нерикардианском условии может выполнять бюджетное ограничение лишь только для некоторых траекторий динамики уровня цен.

Кроме того, они настаивают, что вопрос о наличии права и возможности у государства следовать нерикардианскому условию не имеет смысла для обсуждения, а является лишь предметом веры для конкретного экономиста. Помимо прочего, авторы продемонстрировали пример, основанный на предположениях Вудфорда, в котором возникает результат, несоответствующий интуитивному пониманию действительности: единовременное увеличение денежной массы при нерикардианском условии может привести к гиперинфляции.

Главным критиком Фискальной теории уровня цен можно назвать Уиллема Буайтера. В своей статье (Buiter W., 2002[5]) указывает на проблемы, возникающие на почве ошибок спецификации фундаментальных экономических принципов. Введение нерикардианского условия, по его мнению, приведет к переопределенности системы уравнений в модели, то есть появления множества точек равновесия и, как следствие, возникновению аномальных ситуаций в экономике.

Наборы последовательностей, которыми оперировал Вудфорд в своей работе, определяя, какие из переменных будут экзогенными, а какие эндогенными, а также уравнение (2.18) Буайтер называет фискально-финансово-монетарной программой (FFMP). Под такой программой в трактовке Буайтера понимается полной набор правил, регламентирующий уровень государственных расходов, налогов, трансфертов, выпуска денег и государственных облигаций в каждый момент времени.

Буайтер объясняет, что фискальная теория уровня цен основывается на различии FFMP программ для рикардианского и нерикардианского условий. Нерикардианская FFMP требует, чтобы бюджетное ограничение государства выполнялось только при равновесном уровне цен, но при этом государство всегда обслуживало свой долг.

---
[5] Buiter W., 2002, Ibid.

Рассмотрим модель, которую Буайтер предлагает в своей работе. Это простая динамическая модель общего равновесия с дискретным временем, в которой задействованы репрезентативный агент и государственный сектор. Модель детерминирована, нет никакой неопределенности, а рынки совершенны. Буайтер рассматривает конечное число интервалов времени $1 \leq t \leq N$.

Домашние хозяйства являются ценополучателями на всех рынках, на которых они оперируют. В каждый момент времени они наделяются экзогенным доходом, $y_t > 0$, потребляют $c_t \geq 0$ и платят реальные чистые аккордные налоги $\tau_t$. Помимо этого домохозяйства могут держать три типа активов: наличные деньги (по данному активу не начисляются процентные платежи), номинальные однопериодные облигации с договорной стоимостью $P_t^B \geq 0$, по которым в момент времени $t + 1$ выплачивается номинально измеряемый купон $\Gamma > 0$, и реальные (или индексируемые) однопериодные облигации с договорной ценой $P_t^b \geq 0$, по которым в момент времени $t + 1$ выплачивается измеряемый в реальных величинах купон равный $\gamma > 0$ единиц реального выпуска. Количество денег, номинальных и реальных облигаций автор обозначил как $M_t$, $B_t$ и $b_t$ соответственно, уровень цен в экономике обозначается как $P_t$. Считаем, что государство обладает эксклюзивным правом на выпуск денег.

Обозначим через $i_{t,t+1}$ и $r_{t,t+1}$ однопериодные безрисковые номинальную и реальную ставки процента соответственно. Тогда выполняются следующие соотношения:

$$\mathbf{1 + i_{t,t+1} = \frac{\Gamma}{P_t^B} = \frac{P_{t+1}\gamma}{P_t^b} = (1 + r_{t,t+1})\frac{P_{t+1}}{P_t}}. \qquad (2.21)$$

Назовем эффективными ценами облигаций такие, во сколько их можно оценить, если государство не обслуживает свой долг полностью. Обозначим эффективные цены как $\tilde{P}_t^B$ и $\tilde{P}_t^b$, а символом $D_{t,t+1}$ ту долю долга, которая была погашена государством. Тогда обозначим актуальные платежи по облигациям как

$$\tilde{\Gamma}_{t+1} = D_{t,t+1}\Gamma, \qquad (2.22)$$

$$\tilde{\gamma}_{t+1} = D_{t,t+1}\gamma. \qquad (2.23)$$

Тогда и эффективные цены определяются как

$$\tilde{P}_t^B = D_{t,t+1}P_t^B, \qquad (2.24)$$

$$\tilde{P}_t^b = D_{t,t+1}P_t^b. \qquad (2.25)$$

Буайтер называет $D_{t,t+1}$ фактором переоценки национального долга, а также отмечает, что если $0 \leq D_{t,t+1} < 1$, то этот фактор можно интерпретировать как дефолтный фактор дисконтирования. Отмечаем, что выполняется и соотношение (2.26).

$$1 + i_{t,t+1} = \frac{\tilde{\Gamma}_{t+1}}{\tilde{P}_t^B} = \frac{P_{t+1}\tilde{\gamma}_{t+1}}{\tilde{P}_t^b} = (1 + r_{t,t+1})\frac{P_{t+1}}{P_t}. \qquad (2.26)$$

Запишем однопериодное бюджетное ограничение домашних хозяйств для $1 \leq t \leq N$:

$$M_t - M_{t-1} + \tilde{P}_t^B B_t - D_{t-1,t}\Gamma B_{t-1} + \tilde{P}_t^b b_t - D_{t-1,t}\gamma P_t b_{t-1} \equiv \qquad (2.27)$$

$$\equiv P_t(y_t - c_t - \tau_t),$$

а также наложим ограничение, что домашние хозяйства на конец периода $N$ не могут иметь положительного долга:

$$\widetilde{P}_N^B B_N + \widetilde{P}_N^b b_N \geq 0. \qquad (2.28)$$

Потребители решают свою задачу максимизации функции полезности, заданной выражением (2.29), при ограничениях (2.27) и (2.28) и при заданных начальных значениях активов в их собственности.

$$u_t = \sum_{j=0}^{N-t}\left[\frac{1}{1-\eta}c_{t+j}^{1-\eta} + \phi\frac{1}{1-\eta}\left(\frac{M_{t+j}}{P_{t+j}}\right)^{1-\eta}\right]\left(\frac{1}{1+\delta}\right)^j \qquad (2.29)$$

$$c_{t+j}, M_{t+j} \geq 0, \qquad \eta, \phi, \delta > 0$$

Начальные значения объема активов в портфеле считаем предопределенными:

$$\begin{cases} B_0 = \overline{B}_0, \\ b_0 = \overline{b}_0, \\ M_0 = \overline{M}_0 > 0. \end{cases} \qquad (2.30)$$

Определим номинальный фактор дисконтирования между периодами $t-1$ и $t+j$:

$$R_{t-1,t+j} \equiv \begin{cases} \prod_{k=0}^{j}\frac{1}{1+i_{t-1+k,t+k}}, & j \geq 0 \\ 1, & j = -1 \end{cases} \qquad (2.31)$$

Рекурсивное вычисление (2.27) с подстановкой (2.22) – (2.26) и установка $t = 1$ приводит нас к следующему неравенству:

$$D_{0,1}(\Gamma B_0 + P_1\gamma b_0) \geq \qquad (2.32)$$

$$\geq \sum_{j=0}^{N-1} R_{1,1+j}\bigl[P_{1+j}(c_{1+j} + \tau_{1+j} - y_{1+j}) + (M_{1+j} - M_j)\bigr]$$

Оптимальный выбор домохозяйств характеризуется следующими уравнениями:

$$\left(\frac{c_{t+1}}{c_t}\right)^\eta = (1 + r_{t,t+1})(1+\delta)^{-1}, \quad 1 \le t \le N-1 \qquad (2.33)$$

$$\frac{M_t}{P_t} = c_t\left[\phi\left(\frac{1+i_{t,t+1}}{i_{t,t+1}}\right)\right]^{\frac{1}{\eta}}, \qquad 1 \le t \le N-1 \qquad (2.34)$$

$$\frac{M_N}{P_N} = c_N \phi^{\frac{1}{\eta}}. \qquad (2.35)$$

Уравнение (2.34) дает условие оптимального распределения ресурсов между денежными балансами и потреблением в каждый момент времени, кроме последнего. Уравнение (2.35) указывает на то, что в последний момент времени деньги теряют свою ценность, так как отсутствует какая-либо относительная цена на них.

Государственный сектор представляет собой консолидированные фискальные и монетарные власти. Решения, принимаемые государством, задаются экзогенно, ограничиваясь лишь бюджетным ограничением и условием на погашение своих обязательств. Запишем бюджетное ограничение для $1 \le t \le N$ и условие на долг в конечный момент времени:

$$M_t - M_{t-1} + \widetilde{P}_t^B B_t - D_{t-1,t}\Gamma B_{t-1} + \widetilde{P}_t^b b_t - D_{t-1,t}\gamma P_t b_{t-1} \equiv \qquad (2.36)$$

$$\equiv P_t(g_t - \tau_t),$$

$$\widetilde{P}_N^B B_N + \widetilde{P}_N^b b_N \le 0. \qquad (2.37)$$

Тогда, решая рекурсивно (2.36) и (2.37), получим для начального периода:

$$D_{0,1}(\Gamma B_0 + P_1\gamma b_0) \geq \qquad (2.38)$$

$$\leq \sum_{j=0}^{N-1} R_{1.1+j}[P_{1+j}(\tau_{1+j} - g_{1+j}) + (M_{1+j} - M_j)]$$

Учитывая ограничение домохозяйств (2.32), выражение (2.38) обращается в равенство. Это равенство может выполняться только при совпадении знаков правой и левой части выражения:

$$sgn(\Gamma B_0 + P_1\gamma b_0) = \qquad (2.39)$$

$$= sgn\left\{\sum_{j=0}^{N-1} R_{1.1+j}[P_{1+j}(\tau_{1+j} - g_{1+j}) + (M_{1+j} - M_j)]\right\}$$

Для поиска равновесия задаем основное макроэкономическое тождество и начальные условия (в случае Буайтера – константы в течение всех периодов) для переменных расходов государственного сектора и реального дохода:

$$c_t + g_t = y_t, \qquad 1 \leq t \leq N \qquad (2.40)$$

$$g_t = \bar{g},$$

$$y_t = \bar{y}.$$

По результатам своего исследования Буайтер приходит к заключению, что при нерикардианском условии и экзогенном предложении денег как правила политики, значение уровня цен будет переопределено (более одного значения выявляются в равновесии, которые не обязаны совпадать). При этом он утверждает, что если использовать рикардианское условие с таким же правилом монетарной политики, то уровень цен будет определен однозначно.

При этом Буайтер не смог опровергнуть возможность применения данного подхода, если экзогенно задается процентная ставка. В этом случае он получил детерминированное равновесие.

Но также автор оперирует к аномалиям, возникающим в условиях моделирования при такой трактовке бюджетного ограничения. Обе части уравнения (2.18) ((2.39) соответственно) должны иметь одинаковый знак, однако этого нельзя гарантировать в условиях избыточной экзогенности выбора государством своей макроэкономической политики.

В данной работе будет строиться модель по технологии отличной от методов Буайтера. Будет представлена стохастическая модель с непрерывным временем, и проведена проверка состоятельности критики Буайтера в иных условиях моделирования.

Большинство экономистов, включая авторов и критиков фискальной теории уровня цен, практиковали в своих работах модели с дискретным временем. В случае стохастической динамики, модели с дискретным временем позволяют выявлять равновесные значения только для математических ожиданий исследуемых параметров, при этом теряется возможность учесть дисперсии и ковариации случайных величин. И это может оказаться существенным, если шоки как отклонения от равновесного значения, могут повлиять на само равновесное значение. Особенно это применимо к агентам- рискофобам, например домохозяйствам, чьи функции полезности в доминирующем большинстве случаев задаются вогнутыми.

Чаще всего данный тип моделирования с математическим аппаратом повышенной сложности применяется в теории финансового моделирования и инвестирования. Одними из наиболее известных ученых в данном направлении можно выделить Роберта Мертона (Merton R.C., 1990), Авинаша Диксита и Роберта Пиндайка (Dixit A., Pindyck R., 1994). В России среди крупнейших ученых в данном направлении можно выделить Альберта Ширяева (Shiryaev A., 1999). Стивен Турновски (Turnovsky S., 1995) известен в первую очередь своими работами по стохастической динамике касательно теории экономического роста.

# 3. Модель

## 3.1 Основы стохастических динамических систем

Сформулируем основные понятия непрерывной во времени стохастической динамики.

Назовем *марковским процессом* такой процесс, значения которого с момента времени $t$ и все последующие не зависят от динамики процесса в моменты времени, предшествовавшие $t$.

Назовем *гауссовским процессом* такой процесс, чьи конечномерные распределения гауссовские (нормальные).

Введем понятие *винеровского процесса*, который представляет собой математическую модель непрерывного во времени броуновского движения.

Непрерывный случайный процесс $W_t, t > 0$ назовем *винеровским процессом*, если

1) $W_0 = 0$ почти наверное;
2) $W_t$ является процессом с независимыми приращениями;
3) $W_t - W_z \sim N\left(0, \sigma^2(t-z)\right)$,

где $N(\alpha, \beta)$ – нормальное распределение с математическим ожиданием $\alpha$ и дисперсией $\beta$.

Из определения винеровского процесса следует, что он является гауссовским процессом и марковским процессом.

Свойство (3) можно записать в следующем виде:

$dW = s_t\sqrt{dt}$, где $s_t \sim N(0,1)$ - случайная величина без автокорреляции ($E(s_k s_z) = 0 \ \forall \ k \neq z$). Дисперсия выражается как $VAR(dW) = \sigma_s^2 dt = dt$.

Кроме того, несмотря на то, что винеровский процесс является непрерывным, он нигде не дифференцируем по времени.

В данной работе будут задействованы 2 вида процессов, моделирующих броуновское движение:

- $\frac{dS_1}{S_1} = \mu_1 dt + \sigma_1 dw_1$ - геометрическое броуновское движение,
- $dS_2 = \mu_2 dt + \sigma_2 dw_2$ - броуновское движение с дрейфом.

Так как рассмотренные случайные процессы не дифференцируемы по времени, нужна специальная методика взятия дифференциала функции стохастической переменной. Воспользуемся инструментом, который носит название леммы Ито:

Пусть $F(t, y)$ – функция, зависящая явно от времени и от случайного процесса $y(t)$;

$y(t) = \mu(y,t)dt + dw$;

$dw \sim N(0, \sigma_w^2 dt)$ – стандартное броуновское движение;

пусть существуют производные $\frac{\partial F}{\partial t}, \frac{\partial F}{\partial y}$ и $\frac{\partial^2 F}{\partial y^2}$.

Тогда $dF = \left\{\frac{\partial F}{\partial t} + \frac{\partial F}{\partial y}\mu + \frac{1}{2}\frac{\partial^2 F}{\partial y^2}\sigma_w^2\right\}dt + \frac{\partial F}{\partial y}dw$.

В данной работе мы применим свойство пропорциональности дисперсии интервалу времени для выявления равновесия в FTPL модели, которое выносится в формулирование цели данной главы. Являясь величиной первого порядка, дисперсия не удаляется при рассмотрении бесконечно малых интервалов времени. Равновесие в модели будет определяться как в терминах математических ожиданий, так и в терминах дисперсий и ковариаций. А значит, имеется возможность получить результат моделирования отличный от тех, которые получаются при моделировании в условиях дискретного времени или полной детерминированности.

### 3.2 Общие сведения о динамике модели

Рассмотрим стохастическую модель общего равновесия (задачу динамического программирования) с непрерывным временем. Агентами в данной модели выступают два участника: домохозяйства и консолидация фискальных и монетарных властей в лице единого государственного сектора.

Производственный сектор в модели в явном виде отсутствует. Реальный выпуск задается экзогенно как геометрическое броуновское движение (3.1)

$$\frac{dY}{Y} = g_y^* dt + dy, \quad dy \sim N(0, \sigma_y^2 dt), \qquad (3.1)$$

где $g_y^*$ – ожидаемый (потенциальный) уровень роста логарифма реального ВВП, а $dy$ - стохастический шок выпуска, который мы будем считать нормально распределенным с нулевым математическим ожиданием и дисперсией $\sigma_y^2 dt$.

В модели нет производственного сектора, то есть домохозяйства не могут владеть никаким бизнесом (быть акционерами и получать доход от дивидендов), а также не ходят на работу. Тогда будем считать, что величина $dY$ – это экзогенный доход, которым мы наделяем домохозяйства на отрезок времени $dt$.

Уровень цен в данной модели будет также задаваться геометрическим броуновским движением (3.2)

$$\frac{dP}{P} = \pi^* dt + dp, \quad dp \sim N(0, \sigma_p^2 dt), \qquad (3.2)$$

где $\pi^*$ - ожидаемая скорость роста цен (ожидаемая инфляция), а $dp$ – инфляционный шок, распределенный по нормальному закону с равным нулю математическим ожиданием и дисперсией $\sigma_p^2 dt$.

Сформулируем политику монетарных властей. Она будет иметь пассивный характер и задаваться в форме правила Тейлора:

$$di = [i^* + \beta(\pi^* - \bar{\pi})]dt + \alpha dy. \qquad (3.3)$$

В этой функции номинальная процентная ставка $i$ на момент времени $t$ и до момента времени $t + dt$ устанавливается в зависимости от ожидаемой равновесной ставки $i^*$, отклонения ожидаемой инфляции от целевой $\pi^* - \bar{\pi}$ с параметром $\beta$ и отклонения логарифма реального выпуска от его потенциального уровня $dy$ - шока выпуска – с параметром $\alpha$. При таком определении $di$ отображает номинальную доходность по выраженным в номинальных величинах государственным облигациям на полуинтервале $[t, t + dt)$.

Объявим процесс формирования налоговой политики в виде броуновского движения с дрейфом с ожидаемым установленным уровнем налогов $t$, где вне равновесия величина налогов будет линейно зависеть от экзогенного шока реального дохода населения $dy$.

$$dT = \tau dt + \alpha \frac{B}{P} dy \qquad (3.4)$$

То есть мы формулируем аналог пропорциональной системы налогообложения, весовым коэффициентом в которой является реальный объем номинального долга $B$, помноженный на коэффициент $\alpha$ - тот же, что участвует в правиле Тейлора (3.3). Таким образом, имеет место координация действий монетарных и фискальных властей, когда вторые хотят воспользоваться ростом дохода и увеличить налоги, а первые – компенсируют это предприятие, поднимая в той же степени номинальную ставку доходности. Происходит одновременное ужесточение (или смягчение) фискальной и монетарной политик, направленное на удержание баланса активов и пассивов государственного сектора нечувствительным к шоку выпуска.

Из сформулированного выше выходит, что причиной отклонения значения величины налогов от равновесного уровня является отклонение реального выпуска от потенциального уровня.

Определим реальную стоимость денег как:

$$dR_m \equiv d\left(1/P\right) \Big/ \left(1/P\right). \qquad (3.5)$$

Воспользуемся приближением второго порядка и получим, что $dR_m \cong -\left(\frac{dP}{P}\right) + \left(\frac{dP}{P}\right)^2$. Из того что по определению $\left(\frac{dP}{P}\right)^2 = \sigma_p^2 dt$, выводится реальная стоимость денег:

$$dR_m = \left(-\pi^* + \sigma_p^2\right)dt - dp. \qquad (3.6)$$

По аналогии со стоимостью денег можно вывести функцию реальной доходности по государственным облигациям:

$$dR_B = [i^* - \pi^* + \beta(\pi^* - \overline{\pi}) + \sigma_p^2]dt + \alpha dy - dp, \quad (3.7)$$

что приблизительно соответствует уравнению Фишера, а слагаемое $\sigma_p^2 dt$ можно интерпретировать как премию за инфляционные риски. То есть получаем уравнение:

$$dR_B = di - \frac{dP}{P} + \sigma_p^2 dt. \quad (3.8)$$

Определим номинальное богатство домохозяйств как величину $W$:

$$W = B + M, \quad \frac{M}{W} = \delta, \quad \frac{B}{W} = 1 - \delta, \quad (3.9)$$

где $M$ - объем денежной массы в экономике, $\delta$ - доля всего богатства, которую составляют деньги, а $1 - \delta$ - доля богатства, задействованная под государственные облигации.

Предположим, что всегда существует ненулевая ковариация между относительным приростом выраженных в номинальных величинах государственных обязательств и приростом уровня цен в те моменты времени, когда шок цен имеет место быть (не равен нулю). Если же инфляционный шок отсутствует, и уровень цен находится на равновесной траектории, то очевидно, что прирост государственного долга происходит за счет реального прироста расходов над доходами и корреляция с уровнем цен отсутствует. Домохозяйства, принимая решение об объемах потребления и сбережений, руководствуются ожиданиями об уровне инфляции на следующий период. Тогда объявим дефолт $d\xi$ как снижение значения функции полезности домохозяйств вследствие шока цен и, таким образом, неоптимального распределения располагаемого дохода между потреблением и сбережениями:

$$d\xi = \vartheta \frac{dW}{B} dp, \quad \vartheta = \vartheta(t). \quad (3.10)$$

$d\xi$ характеризует снижение инвестиционной привлекательности облигаций и денег. Для государства в терминах данной модели появление ненулевого композиционного шока $d\xi$ может представлять угрозу нарастить свои обязательства в следующий момент времени $t + dt$ в меньшем объеме, чем это требует образовавшийся дефицит бюджета, из-за неучтенного ценового шока, так как в сценарий запланированного бюджета закладывается ожидаемая инфляция. И это приводит правительство к "дефолту" – оно не признает долгом часть образованного дефицита. Технически это равносильно доли накопленного на текущий момент времени долга, по которому государство объявляет дефолт, то есть отказывается его обслуживать.

С другой стороны, отрицательные значения величины $d\xi$ приведут к избыточности наращивания государственных обязательств, в том числе и долговых. При экзогенном задании последовательности государственных расходов и налогов, которые не компенсируют любой шок государственных закупок, это может привести к неустойчивости системы обслуживания обязательств. В итоге снова возникает вероятность дефолта.

Очевидно, что номинальные обязательства государственного сектора всегда положительно коррелируют с уровнем цен (если эта корреляция существует), а знак функции $\vartheta$ совпадает со знаком шока цен. Причем функция $\vartheta$ положительна при нулевом шоке. Значит, индикатор $d\xi$ и величина дефолта может принимать как положительные, так и отрицательные значения. В ходе решения модели (поиска равновесного состояния) величина $d\xi$ будет играть роль индикатора того, что система не в равновесии, и имеется ненулевой шок $dp$, отталкивающий систему от равновесной траектории.

**3.3 Домохозяйства**

Поведение домохозяйств можно представить функцией полезности следующего вида

$$U(C, m) = \frac{1}{\gamma}\left(C^\theta m^{1-\theta}\right)^\gamma, \quad (3.11)$$

в которой описываются предпочтения между величиной реального потребления $C$ и реальных денежных балансов $m$. Задачу домохозяйств можно записать как максимизацию суммы дисконтированных значений функции полезности за весь исследуемый временной полуинтервал $[0, +\infty)$:

$$\max_{C,\delta} \int_0^{+\infty} U(C, m) e^{-\rho t} dt \quad (3.12)$$

при бюджетном ограничении

$$dw = dY - Cdt - dT + dR_m \frac{M}{P} + dR_B \frac{B}{P} - d\xi \frac{B}{P}. \quad (3.13)$$

Выразим реальный объем активов

$$w = W/P \quad (3.14)$$

и подставим (1) – (10) в (13):

$$dw = Y g_y^* dt + Y dy - Cdt - \tau dt - \alpha \frac{B}{P} dy + w(-\pi^* + \sigma_P^2)dt - wdp + \quad (3.15)$$
$$+ (1-\delta)w[i^* + \beta(\pi^* - \bar{\pi})]dt + \alpha(1-\delta)wdy - \vartheta dW\, dp$$

Произведение стохастических процессов дает ковариацию между ними:

$$dW * dp = \sigma_{WP} dt \quad (3.16)$$

Приращение реальных активов можно расписать как сумму детерминированной и стохастической компонент:

$$dw = fdt + d\lambda \quad (3.17)$$

$$f = Y g_y^* - C - \tau - \vartheta \frac{\sigma_{WP}}{P} + w\left[-\pi^* + \sigma_p^2 + (1-\delta)(i^* + \beta(\pi^* - \bar{\pi}))\right] \quad (3.18)$$

$$d\lambda = Ydy - wdp \qquad (3.19)$$

$d\lambda$ является стохастической компонентой, распределенной по нормальному закону с математическим ожиданием равным нулю и дисперсией $\sigma_\lambda^2 dt$, где

$$\sigma_\lambda^2 = Y^2\sigma_y^2 - 2Yw\sigma_{yp} + w^2\sigma_p^2. \qquad (3.20)$$

Пусть $X(w)$ – функция ценности, тогда должно выполняться

$$\max_{C,\delta}\left\{U(C,\delta) + \frac{\partial X}{\partial t} - \rho X + f*X_w + \frac{1}{2}\sigma_\lambda^2 X_{ww}\right\} = 0 \qquad (3.21)$$

Выпишем частные производные:

$$f'_\delta = -w[i^* + \beta(\pi^* - \bar{\pi})] \qquad (3.22)$$

$$U'_C = \theta(C^\theta m^{1-\theta})^{\gamma-1}\left(\frac{C}{w}\right)^{\theta-1}\delta^{1-\theta} \qquad (3.23)$$

$$U'_\delta = (1-\theta)(C^\theta m^{1-\theta})^{\gamma-1}\left(\frac{C}{w}\right)^\theta w\delta^{-\theta} \qquad (3.24)$$

Условия первого порядка:

$$U'_C = X_w \qquad (3.25)$$

$$U'_\delta - w(i^* + \beta(\pi^* - \bar{\pi}))X_w = 0 \qquad (3.26)$$

Пусть $X(w)$ имеет следующий вид:

$$X_{ww} = \frac{\theta\gamma-1}{w}X_w; \quad X_w = \tilde{A}\theta\gamma w^{\theta\gamma-1}; \quad X = \tilde{A}w^{\theta\gamma}, \qquad (3.27)$$

тогда

$$\frac{U'_\delta}{U'_C} = w(i^* + \beta(\pi^* - \bar{\pi})). \qquad (3.28)$$

Условие (3.28) дает нам условие, выражающее предельную норму замещения:

$$\frac{U'_m}{U'_C} \equiv MRS_{mc} = i^* + \beta(\pi^* - \bar{\pi}), \qquad (3.28.1)$$

так как

$$\frac{U'_\delta}{U'_m} = w.$$

Далее из (3.23), (3.24) и (3.28) следует

$$\frac{1-\theta}{\theta}\frac{C}{w}\frac{1}{\delta} = \left(i^* + \beta(\pi^* - \overline{\pi})\right), \tag{3.29}$$

$$\frac{C}{w} = \frac{\theta\delta}{1-\theta}\left(i^* + \beta(\pi^* - \overline{\pi})\right). \tag{3.30}$$

Подставляем (3.23), (3.27) и (3.30) в (3.25):

$$(\delta w)^{(1-\theta)\gamma} = \widetilde{A}\gamma\left(\frac{C}{w}\right)^{1-\theta\gamma}, \tag{3.31}$$

$$\delta^{\gamma-1} = \widetilde{A}\frac{\gamma}{w^{(1-\theta)\gamma}}\left[\frac{\theta}{1-\theta}[i^* + \beta(\pi^* - \overline{\pi})]\right]^{1-\theta\gamma} \tag{3.32}$$

Далее подставляем полученные промежуточные результаты в функцию Беллмана (3.21):

$$\frac{1}{\gamma}\left(\left[\frac{\theta}{1-\theta}[i^* + \beta(\pi^* - \overline{\pi})]\right]^\theta w\delta\right)^\gamma - \rho\widetilde{A}w^{\theta\gamma} +$$

$$+ \widetilde{A}\theta\gamma w^{\theta\gamma-1}\left(Yg_y^* - C - \tau - \vartheta\frac{\sigma_{WP}}{P} + w[-\pi^* + \sigma_P^2 + (1-\delta)(i^* + \beta(\pi^* - \overline{\pi}))]\right) +$$

$$+ \frac{1}{2}\left[(\theta\gamma - 1)\widetilde{A}\theta\gamma w^{\theta\gamma-2}\sigma_\lambda^2\right] = 0.$$

Выражаем $\widetilde{A}$ как функцию от $\delta$:

$$\widetilde{A} = \frac{\left[\frac{\theta}{1-\theta}[i^* + \beta(\pi^* - \overline{\pi})]\right]^{\theta\gamma}\delta^\gamma w^{\gamma(1-\theta)}}{\gamma\left\{\rho - \theta\gamma\left[\frac{Y}{w}g_y^* + \left(1 - \frac{\delta}{1-\theta}\right)[i^* + \beta(\pi^* - \overline{\pi})] + \frac{(\theta\gamma-1)}{2w^2}\sigma_\lambda^2 - \frac{\tau}{w} - \vartheta\frac{\sigma_{WP}}{wP} - \pi^* + \sigma_P^2\right]\right\}}$$

И подставим полученный результат в выражение (3.32):

$$\delta^{-1} = \frac{\dfrac{\theta}{1-\theta}[i^* + \beta(\pi^* - \overline{\pi})]}{\rho - \theta\gamma\left[\dfrac{Y}{w}g_y^* + \left(1 - \dfrac{\delta}{1-\theta}\right)[i^* + \beta(\pi^* - \overline{\pi})] + \dfrac{(\theta\gamma - 1)}{2w^2}\sigma_\lambda^2 - \dfrac{\tau}{w} - \vartheta\dfrac{\sigma_{WP}}{wP} - \pi^* + \sigma_p^2\right]}$$

Таким образом, мы находим оптимальную долю денежных активов $m$ среди всех сбережений $w$:

$$\boldsymbol{\delta = \frac{(1-\theta)}{\theta(1-\gamma)} *} \qquad (3.33)$$

$$* \frac{\rho - \theta\gamma\left[\dfrac{Y}{w}g_y^* - \dfrac{\tau}{w} - \vartheta\dfrac{\sigma_{WP}}{wP} + i^* + \beta(\pi^* - \overline{\pi}) - \pi^* + \sigma_p^2 + \dfrac{(\theta\gamma - 1)}{2w^2}\sigma_\lambda^2\right]}{i^* + \beta(\pi^* - \overline{\pi})}$$

И подстановка (3.33) в (3.30) дает нам условие оптимального выбора домохозяйств:

$$\frac{C}{w} = \frac{1}{1-\gamma} * \qquad (3.34)$$

$$* \left\{\rho - \theta\gamma\left[\frac{Y}{w}g_y^* - \frac{\tau}{w} - \vartheta\frac{\sigma_{WP}}{wP} + i^* + \beta(\pi^* - \overline{\pi}) - \pi^* + \sigma_p^2 + \frac{(\theta\gamma - 1)}{2w^2}\sigma_\lambda^2\right]\right\}$$

Заметим, что домашние хозяйства, принимая решения об отношении объема потребления к объему инвестиций, формируют свои планы одинаково для любого уровня цен, а значит, не реагируют на его изменение. Тогда выбор, который совершают домохозяйства, описывается уравнениями (33.1) и (34.1).

$$\delta = \frac{(1-\theta)}{\theta(1-\gamma)} * \quad (3.33.1)$$

$$* \frac{\rho - \theta\gamma\left[\frac{Y}{w}g_y^* - \frac{\tau}{w} + i^* + \beta(\pi^* - \bar{\pi}) - \pi^* + \sigma_p^2 + \frac{(\theta\gamma-1)}{2w^2}\sigma_\lambda^2\right]}{i^* + \beta(\pi^* - \bar{\pi})}$$

$$\frac{C}{w} = \frac{1}{1-\gamma} * \quad (3.34.1)$$

$$* \left\{\rho - \theta\gamma\left[\frac{Y}{w}g_y^* - \frac{\tau}{w} + i^* + \beta(\pi^* - \bar{\pi}) - \pi^* + \sigma_p^2 + \frac{(\theta\gamma-1)}{2w^2}\sigma_\lambda^2\right]\right\}$$

Уравнения (33), (34), (33.1) и (34.1) позволяют нам сформулировать <u>необходимое условие стабильности уровня цен</u>: *если цены и инфляция находятся на равновесном уровне (инфляционный шок равен нулю), то оптимальный выбор домохозяйств описывается уравнениями (3.33.1) и (3.34.1).* Если в системе присутствует ценовой шок, домохозяйства не смогут максимизировать значение функции полезности, совершая выбор по данному правилу.

Данное условие можно сформулировать и в терминах теории игр. Модель можно представить как игру с последовательными действиями, где государственный сектор является лидером, а домохозяйства – ведомым агентом. Тогда необходимое условие звучит следующим образом: *если цены и инфляция находятся на равновесном уровне, то уравнениям (3.33.1) и (3.34.1) описывают лучший ответ домохозяйств.*

## 3.4 Государственный сектор

Далее рассмотрим государственный сектор. Сформулируем функцию государственных расходов как их приращение относительно выпуска. Будем считать, что доля государственных расходов в общем выпуске на полуинтервале $[t, t + dt)$ определяется средним уровнем $s_g^* dt$ и линейной нормально распределенной стохастической компонентой $dg$ с нулевым математическим ожиданием и дисперсией $\sigma_g^2 dt$:

$$\frac{dG}{Y} = s_g^* dt + dg, \quad dg \sim N(0, \sigma_g^2 dt). \tag{3.35}$$

Таким образом мы задаем свободу действий государства в формировании собственных расходов.

Предложение денег определяем геометрическим броуновским движением, в котором объем денежной массы растет со средней скоростью $\mu$ и случайной компонентой $du$:

$$\frac{dM}{M} = \mu dt + du, \quad dg \sim N(0, \sigma_u^2 dt). \tag{3.36}$$

Из (3.2) и (3.36) можно выявить динамику реальных денежных балансов $m$:

$$m = \frac{M}{P}, \quad \frac{dm}{m} = (\mu - \pi^* - \sigma_{up}) + du - dp. \tag{3.37}$$

В то время, когда монетарные власти устанавливают процентную ставку в соответствии с правилом Тейлора, домохозяйства принимают решение об объеме спроса на денежные балансы. И после этого монетарные власти готовы предложить ровно столько денег экономике, сколько необходимо для достижения равновесия на денежном рынке.

Запишем бюджетное ограничение государственного сектора в виде (3.38)

$$dB + dM = d(PG) - d(PT) + Bdi - Bd\xi, \tag{3.38}$$

где слева зафиксировано обеспечение дефицита бюджета через выпуск новых облигаций и увеличение денежной массы, а справа – образованный дефицит – разность государственных расходов (государственные закупки и платежи по накопленному долгу) и доходов (налоги) с вычетом той части долга, по которой может быть объявлен дефолт (в терминах модели) из-за неучтенного ценового шока.

Мы строим модель экономики с активной фискальной политикой, где правительство независимо формирует свой бюджет, выбирая уровни государственных расходов и налогов на каждый период времени, и таким образом сообщает о своих текущих и будущих излишках и дефиците. Монетарная политика принимает как данное бюджетное ограничение, и может попытаться профинансировать дефицит через наращивание денежной массы. Описание активной фискальной политики сформулировано у Сарджента и Уоллеса (Sargent, T., Wallace, N., 1981[6]), название предложено Липером (Leeper, E.M., 1991[7]).

Подставим (3.1)-(3.4), (3.10) и (3.35) в (3.38):

$$dW = [P(Ys_g^* - \tau + Y\sigma_{gp} - \sigma_{yp} + (G-T)\pi^*) + B(i^* + \beta(\pi^* - \bar{\pi}))]dt + \quad (3.39)$$
$$+ P[Ydg + (G-T)dp] - \vartheta dWdp$$

Здесь $\sigma_{gp}dt$, $\sigma_{yp}dt$ – ковариации реальных государственных расходов и уровня цен, а также реального выпуска и уровня цен соответственно.

$$E(dy) = E(dp) = 0 \quad (3.40)$$

Воспользуемся условием отсутствия "пирамид" (No-Ponzi Game condition)

---

[6] Thomas J. Sargent, Neil Wallace, Fall 1981, Ibid. P.2.
[7] Eric M. Leeper. Equilibria under 'active' and 'passive' monetary and fiscal policies. // Journal of Monetary Economics 27, P. 129-147, 1991.

$$\lim_{t\to\infty} E_t\left[W(t)e^{-(i^*+\beta(\pi^*-\overline{\pi}))t}\right] = 0. \tag{3.41}$$

Проинтегрируем левую и правую части уравнения (39) и выведем требуемую величину государственных обязательств:

$$\frac{W(t)+E_t\int_t^\infty \vartheta\sigma_{WP}e^{-(i^*+\beta(\pi^*-\overline{\pi}))(z-t)}dz}{P(t)} = \tag{3.42}$$

$$= E_t\int_t^\infty \frac{P_z}{P_t}\left[\left(-i^* + \frac{1-\delta}{\delta}\beta(\pi^*-\overline{\pi})\right)m + Ys_g^* - \tau + Y\sigma_{gp} - \sigma_{yp} + (G-T)\pi^*\right] *$$

$$e^{-(i^*+\beta(\pi^*-\overline{\pi}))(z-t)}dz$$

Принимая во внимание то, что $E\frac{dP}{P} = \pi^*dt$, $E\frac{P_z}{P_t} = e^{\pi^*(z-t)}$, приходим к (3.43):

$$\frac{W(t)}{P(t)} = E_t\int_t^\infty \left[\left(-i^* + \frac{1-\delta}{\delta}\beta(\pi^*-\overline{\pi})\right)m + Ys_g^* - \tau + Y\sigma_{gp} - \sigma_{yp} + (G-T)\pi^*\right] * \tag{3.43}$$

$$e^{-(i^*-(\beta\overline{\pi}+(1-\beta)\pi^*))(z-t)}dz -$$

$$-E_t\int_t^\infty \vartheta\sigma_{WP}e^{-(i^*-(\beta\overline{\pi}+(1-\beta)\pi^*))(z-t)}dz$$

Фактически (3.43) является функцией, описывающей предложение государственных обязательств. Последнее слагаемое в равновесии обнуляется, так как шок цен на всей траектории равен нулю, согласно определению (3.10). Этот интеграл характеризует суммарный объем реальных единиц, которые домохозяйства излишне инвестировали в государство, не учтя ценовой шок. В результате этого значение функции полезности отклоняется от максимального значения. С точки зрения государства это суммарный дефолт в текущих ценах.

## 3.5 Равновесие

Определим 7 ключевых уравнений, которые определяют равновесие в модели:

$$C = \frac{\theta\delta}{1-\theta}\frac{W}{P}\bigl(i^* + \beta(\pi^* - \overline{\pi})\bigr), \qquad (3.44)$$

$$\delta = \frac{(1-\theta)}{\theta(1-\gamma)} * \qquad (3.45)$$

$$* \frac{\rho - \theta\gamma\left[\frac{Y}{W}g_y^* - \frac{\tau}{W} + i^* + \beta(\pi^* - \overline{\pi}) - \pi^* + \sigma_p^2 + \frac{(\theta\gamma - 1)}{2w^2}\sigma_\lambda^2\right]}{i^* + \beta(\pi^* - \overline{\pi})}$$

$$\frac{dP}{P} = \pi^* dt + dp, \qquad (3.46)$$

$$\frac{dM}{M} = \mu dt + du, \qquad (3.47)$$

$$M = \delta W, \qquad (3.48)$$

$$dY = Cdt + dG, \qquad (3.49)$$

$$\frac{W}{P} = \Phi. \qquad (3.50)$$

Здесь уравнения (3.44) – (3.45) аналогичны уравнениям (3.30) и (3.33.1) соответственно, уравнения (3.46), (3.47) соответствуют определениям (3.2) и (3.36), величина денежной массы в (3.48) определяется в соответствии с (3.9), уравнение (3.49) – основное макроэкономическое тождество с отсутствием инвестиций в закрытой экономике, а уравнение (3.50) релевантно уравнению (3.43), где мы через Ф обозначим выражение

$$E_t \int_t^\infty \left[\left(-i^* + \frac{1-\delta}{\delta}\beta(\pi^* - \overline{\pi})\right)m + Ys_g^* - \tau + Y\sigma_{gp} - \sigma_{yp} + (G-T)\pi^*\right] *$$

$$e^{-(i^* - (\beta\overline{\pi} + (1-\beta)\pi^*))(z-t)}dz -$$

$$-\int_t^\infty \vartheta\sigma_{WP} e^{-(i^*-(\beta\bar{\pi}+(1-\beta)\pi^*))(z-t)} dz.$$

Уравнения (3.46) – (3.47) являются определениями динамик роста цен и денежной массы, а уравнения (3.44) – (3.45), (3.48) – (3.50) определяют двойные условия для переменных в терминах математических ожиданий, дисперсий и ковариаций и являются двойными условиями.

В модели присутствуют 12 эндогенных переменных: $\delta, C, W, P, \pi^*, \mu,$ $\sigma_p^2, \sigma_u^2, \sigma_{gp}, \sigma_{up}, \sigma_{yp}$ и комплексная переменная $\vartheta\sigma_{WP}$. А уравнения (3.44) – (3.50) определяют 12 условий, позволяющих точно идентифицировать эндогенные переменные единственным способом.

Значение равновесной номинальной процентной ставки $i^*$ может быть выявлено из условия (3.28.1) о предельной норме замещения, а также выбора экзогенных переменных $\beta, \bar{\pi}$ монетарными властями и определения эндогенной переменной $\pi^*$. А величина реальных обязательств государства (реального богатства) $w$ и ее дисперсия $\sigma_\lambda^2$ выявляются из условий со значениями эндогенных переменных (3.14) и (3.20).

Таким образом, была предъявлена модель, в которой для всех эндогенных переменных можно предъявить однозначные решения, а равновесие может быть достигнуто таким выбором экзогенных переменных, при которых переменная $\vartheta\sigma_{WP}$ обращается в ноль на всей своей траектории.

## 4. Заключение

Целью данной работы было исследование равновесия в модели фискальной теории идентификации уровня цен на предмет переопределенности в условиях стохастического моделирования в непрерывном времени. Критики данной теории высказывают свои аргументы, основываясь на моделях с дискретным временем. Основной задачей данной работы было проверить состоятельность этой критики, то есть предъявить такую модель, в которой озвученные ранее аргументы против данной теории нецелесообразны.

Ключевым свойством стохастического моделирования в непрерывном времени является возможность сформулировать условия равновесия как в терминах математических ожиданий, так и дисперсиях и ковариациях. Таким образом, число уравнений в системе, определяющей равновесие, отличается от результатов моделирования в дискретном времени и в условиях полной детерминированности.

В работе построена модель общего равновесия, которая предусматривает стохастическую динамику в непрерывном времени, где имеется стохастически описанная возможность дефолта, привязанная к шоку уровня цен. Монетарные власти следуют правилу номинальной процентной ставки, представленному в виде правила Тейлора. Было предъявлено необходимое условие стабильности уровня цен для действий домохозяйств и выявлена система уравнений, описывающих равновесие. Показано, что в данной модели равновесие не является переопределенным, что противоречит выводам критиков. Значит, результаты данной работы можно отнести к защите основных положений FTPL.

Прошло уже более 20 лет с момента написания первых трудов, посвященной фискальной теории уровня цен, но она все еще остается довольно новым открытием в макроэкономической науке. Не вызывает удивления то, что многие могут с ней не соглашаться критиковать, и это лишь служит поводом для дальнейших исследований в данной области.

Результаты данной работы показывают, что нельзя отказаться от какого-либо предположения только лишь потому, что она не вписывается в рамки отдельно взятой модели. Необходимо понимать, какие теоретические предпосылки и методы моделирования в каком случае более точно описывают реальную жизнь, и пытаться проводить исследования уже в контексте этих условий.

Модель в данной работе не претендует на звание единственно верной. Более того, она является существенно упрощенной и в последующих исследованиях нуждается в дополнительных уточнениях. В первую очередь необходимо заменить предположение об экзогенном доходе и ввести определение производственного сектора. В отсутствие реального сектора реальная ставка процента определяется через номинальную ставку и инфляцию. Кроме того, имеется смысл ввести отдельно сформулированный финансовый сектор.

Помимо тестирований в рамках моделей, надо проводить исследования с историческими данными реальных экономик и пытаться выявить, как политика фискальных властей в действительности влияла на обеспечение стабильности уровня цен.

# Список литературы

1. Alesina A., Tabellini G. (1987). Rules and Discretion with Noncoordinated Monetary and Fiscal Policies. Economic Inquiry, 12. P. 619-630.

2. Beetsma R.M.W.J., Bovenberg L. (1995). The Role of Public Debt in the Double Game of Chicken. Mimeo.

3. Beetsma R.M.W.J., Bovenberg L. (1997). Designing Fiscal and Monetary Institutions in a Second-Best World. European Journal of Political Economy, 13. P. 53-79.

4. Beetsma R.M.W.J., Bovenberg L. (1997). Central Bank Independence and Public Debt Policy. Journal of Economic Dynamics and Control, 21, P. 873-894.

5. Blinder A. (1982). Issues in the Coordination of Monetary and Fiscal Policy. NBER Working Paper No. 982.

6. Buiter W.H. (2002). The Fiscal Theory of the Price Level: A Critique. The Economic Journal, 112. P. 459-480.

7. Christ C.F. (1979). On Fiscal and Monetary Policies and Government Budget Restraint. American Economic Review, Vol. 69, No. 4. P. 526-538.

8. Cochrane J.H. (2000). Money as Stock: Price Level Determination with No Money Demand. NBER Working Paper No. 7498.

9. Cochrane J.H. (2001). Long-Term Debt and Optimal Policy in the Fiscal Theory of the Price Level. Econometrica, 69. P. 69-116

10. Cochrane J.H. (2003). Fiscal Foundations of Monetary Regimes.

11. Dixit A.K., Pindyck R.C. (1994). Investment under Uncertainty. Princeton University Press.

12. Fischer S., Sahay R., Vegh C.A. (1994). Modern Hyper- and High Inflations. Journal of Economic Literature, 40 (3). P. 837-880.

13. Fisher W.H., Turnovsky S.J. (1992). Fiscal Policy and the Term Structure of Interest Rates: An Intertemporal Optimizing Analysis. Journal of Money, Credit and Banking. Vol. 24, No. 1. P. 1-26.

14. Grinols E.L., Turnovsky S.J. (1998). Risk, Optimal Government Finance and Monetary Policies in a Growing Economy. Econometrica, 65. P. 401-427.

15. Giuliano P., Turnovsky S. (2000). Intertemporal Substitution, Risk Aversion, and Economic Performance in a Stochastically Growing Open Economy.

16. King R.G., Plosser C.I. (1985). Money, Deficits, and Inflation. Carnegie-Rochester Conference on Public Policy, 22. P. 191-222.

17. Kocherlakota N., Phelan C. (1999). Explaining the Fiscal Theory of the Price Level. Federal Reserve Bank of Minneapolis Quarterly Review, 23 (4). P. 14-23

18. Leeper E. (1991). Equilibria Under "Active" and "Passive" Monetary and Fiscal Policies. Journal of Monetary Economics, 27. P. 129-147

19. Leeper E.M., Yun T. (2005). Monetary-Fiscal Policy Interactions and the Price Level: Background and Beyond. NBER Working Paper No. 11646.

20. Lucas R.E. Jr., Stokey N.L. (1983). Optimal fiscal and monetary policy in an economy without capital. Journal of Monetary Economics, Vol.12 (1). P. 55-93.

21. Merton R.C. (1969). Lifetime Portfolio Selection Under Uncertainty: The Continuous-Time Case. Review of Economics and Statistics, 51. P. 247-25. Reprinted in Merton R.C. Continuous-time Finance, 1990, as Chapter 4.

22. Merton R.C. (1971). Optimum Consumption and Portfolio Rules in a Continuous-Time Model. Journal of Economic Theory, 3. P. 373-413. Reprinted in Merton R.C. Continuous-time Finance, 1990, as Chapter 5.

23. Merton R.C. (1973). An Intertemporal Capital Asset Pricing Model. Econometrica, 41. P. 86-887. Reprinted in Merton R.C. Continuous-time Finance, 1990, as Chapter 15.

24. Merton R.C. (1975). Theory of Finance from the Perspective of Continuous Time. Journal of Financial and Quantitative Analysis, 10. P. 659-674.

25. Merton R.C. (1990). Continuous-time Finance. Basil Blackwell.

26. Sargent T.J. (1985). Reaganomics and Credibility. In: A. Ando et al. eds. Monetary Policy. MIT Press: Cambridge. (Reprinted in: Sargent T.J. Rational Expectations and Inflation. 2$^{nd}$ Ed. 1993. Harper Collins College Publishers: New York.)

27. Sargent T.J., Wallace N. (1975). Rational Expectations, the Optimal Monetary Instrument, and the Optimal Money Supply Rule. Journal of Political Economy, Vol. 83, No. 2. P. 241-254.

28. Sargent T.J., Wallace N. (1981). Some Unpleasant Monetarist Arithmetic. Federal Reserve Bank of Minneapolis Quarterly Review. P. 1-17.

29. Shiryaev A.N. (1999). Essentials of Stochastic Finance: Facts, Models, Theory. World Scientific, Advanced Series on Statistical Science and Applied Probability, Vol. 3.

30. Sidrauski M. (1967). Rational Choice and Patterns of Growth in a Monetary Economy. American Economic Review, 57 (2). P. 534-544.

31. Sims C.A. (1994). A Simple Model for the Determination of the Price Level and the Interaction of Monetary and Fiscal Policy. Economic Theory, 4. P. 381-399.

32. Turnovsky S. (1995). Methods of macroeconomic dynamics. Massachusetts Institute of Technology.

33. Woodford M. (May 1994). Monetary policy and price level determinacy in a cash-in-advance economy. Economic Theory, Vol.4 (3). P. 345-380

34. Woodford M. (1995). Price-level Determinacy without Control of a Monetary Aggregate. Carnegie-Rochester Conference Series on Public Policy 43. P. 1-46.

35. Woodford M. (1996). Control on the Public Debt: A Requirement for Price Stability? NBER Working Paper No. 5684.

36. Woodford M. (1998). Public Debt and the Price Level. Discussion Paper, Princeton University.

37. Woodford M. (2001). Fiscal Requirements for Price Stability. NBER Working Paper No. 8072.